\documentclass[12pt]{article}
\usepackage{epsfig}
\include{graphics}

\include{epsfig}
\begin{document}
\begin{center}
{\Large {\bf Green functions in graphene monolayer with Coulomb interactions
taken into account}}

\vskip-30mm \rightline{\small ITEP-LAT/2011-05 } \vskip 30mm

{%\baselineskip=16pt
\vspace{1cm}
{M.V. Ulybyshev$^{a,b,}$\footnote{e-mail: ulybyshev@goa.bog.msu.ru} }, { M.A.~Zubkov$^{b,}$\footnote{e-mail: zubkov@itep.ru} }\\
\vspace{.5cm} {\it $^a$ Institute for Theoretical Problems of Microphysics,
Moscow State
University, Moscow, Russia }\\
\vspace{.5cm} {\it $^b$ ITEP, B.Cheremushkinskaya 25, Moscow, 117259, Russia }
}

\end{center}

\begin{abstract}
 We consider the low energy effective field model of graphene monolayer. Coulomb interactions are taken into account. The model is simulated numerically using the lattice discretization with staggered fermions. The two point fermionic Green functions are calculated. We find that in the insulator phase these Green functions almost do not depend on energy. This indicates that the effective field model (in its insulator phase) does not
correspond to the real graphene. 
\end{abstract}

%\documentclass[12pt]{article}

%%%%%%%%%%%%%%%%%%%%

\newcommand{\br}{{\bf r}}
\newcommand{\bu}{{\bf \delta}}
\newcommand{\bk}{{\bf k}}
\newcommand{\bq}{{\bf q}}
\def\({\left(}
\def\){\right)}
\def\[{\left[}
\def\]{\right]}

\section{Introduction}

It is well - known that without the Coulomb interactions the effective field
model of graphene monolayer is a good approximation to the original tight -
binding model. This effective field model \cite{Katsbook,Shytov,bi,Novoselov:04:1,Semenoff,Fialkovsky,Geim:07:1,Wallace,CastroNeto:2009zz}
operates with the continuum Dirac field living in the graphene sheet.  This
continuum model is used also when the Coulomb interactions are switched on\footnote{In this case the continuum Dirac field interacts with the dynamical
field of the electric potential.}. We suppose that it remains a good
approximation to the tight - binding model when this effective field model
remains in the semi - metal phase, i.e. it does not predict the appearance of
the energy gap. However, as it will be explained below, we have some doubts
that this model may be applicable for the small values of the substrate
dielectric permittivity, where it predicts the appearance of the fermion
condensate (see below).

 Recently, the effective field model of graphene monolayer
with the Coulomb interactions taken into account was investigated numerically using nonperturbative lattice methods \footnote{Within the ranges of perturbation theory the effect of the Coulomb interaction on various physical
quantities was investigated in a number of papers (see, for example,  \cite{perturb1, perturb2} and references therein). }. The
application of numerical lattice methods is justified by the fact that the
Fermi velocity $v_F$ is about $1/300$. That's why the effective coupling
constant $\alpha \sim \frac{1}{137 v_F}$ is large and the Coulomb interactions
are strong. Therefore, nonperturbative effects may be strong.  In
\cite{Araki,Araki:10:1,Timo,Itep2012,armour,Drut:10:1,Hands:11:1,Lahde:09:1,Lahde:09:3,Lahde:11:1}
the effective low energy field model of graphene was investigated numerically
using the lattice regularized model with staggered fermions \footnote{Within
the original tight - binding model the problem was considered analytically in
\cite{Araki:12:1} while in \cite{Rebbi} it was investigated numerically.}.  The
main output of these investigations is that there exists the phase transition
at a certain value of the effective coupling constant $\beta$. This effective
coupling constant is related to the dielectric permittivity $\epsilon$ of
substrate as follows \cite{Itep2012}:
\begin{equation}
\beta \approx \frac{137}{300} \frac{1}{4 \pi} \frac{\epsilon+1}{2}.
\label{beta_def}
\end{equation}
(It is worth mentioning that due to the lattice artifacts this relation may be modified within a certain lattice realization of the effective field model of graphene. )

 There
is an evidence that this is the semi - metal -- insulator phase transition.
Namely, one of the possible condensates becomes nonzero at $\beta < \beta_c$
\cite{Timo,Itep2012}. In addition, the indications were found that the usual
longitudinal conductivity vanishes at the position of the phase transition
\cite{Itep2012}. The possibility that the insulator phase may appear in graphene monolayer has also been discussed in another context (see, for example, \cite{gu} and references therein).

The possibility that the effective low energy field model describes well the real graphene
 is not so obvious when the effective
low energy model is in the insulator phase. Our conclusions are based on the direct measurement of the two - point
Green function in the lattice regularized effective field model of graphene.
The regularization is based on staggered fermions. We simulate the model using
the same code that was used earlier by one of us during the work on the paper
\cite{Itep2012}. This code was tested in several ways (in particular, some
previous results on the graphene monolayer
\cite{Timo,Lahde:09:1,Lahde:09:3,Lahde:11:1} were reproduced). We demonstrate
that in the insulator phase the Green function almost does not depend on energy
while its dependence on the space - like momentum remains nontrivial. This
means that the correlation time  becomes negligible.
 At the same time the correlation length in physical units may remain nonzero or, even, infinite. Therefore,
the physical energy of the fermion excitation tends to infinity and
 the given field - theoretical model is not self - consistent at the
corresponding values of $\epsilon$ and, thus, cannot describe the real physics.

The paper is organized as follows. In Section \ref{SectModel} we consider the
details of the model.  In Section \ref{SectNum} numerical results are
represented. In Section \ref{SectConcl} we end with the conclusions.

\section{The effective field model of graphene monolayer in lattice regularization}
\label{SectModel}

%\subsection{Green functions expressed through two - component spinors}

In the present paper we  use notations adopted in \cite{Z2011} and \cite{Itep2012}. The model
contains two flavors (corresponding to spin) of the $4$ - component spinors $\psi$
coupled to the electric potential $A_4$ (we work in the imaginary time representation). The Green function has to be considered in a certain gauge. The gauge freedom of the system corresponds to the transformation $ A_4
\rightarrow A_4 +
\partial_4 \alpha(x^4) \quad \psi \rightarrow e^{i\alpha} \psi $. In our
numerical procedure we fix this gauge freedom via the condition $A_4(x^4, {\bf
z}) = 0$ for the  $3D$ point $\bf z = 0$.
(We unfix the value of $A_4$ at a certain point on
this line.) The two - point fermion Green function has the form:
\begin{eqnarray}
&& {\bf G}  =  \langle {\psi}^{\dag}_x \psi_{y} \rangle   = \frac{1}{Z}\int
D\bar{\psi}D\psi D A {\psi}^{\dag}_x \psi_{y}{\rm exp} \Bigl( - \frac{1}{2}\int
d^4x [\partial_{I} A_{4}]^2 \nonumber\\&& - \int d^3x \bar{\psi}([\partial_4 -
i g A_4] \Gamma^4 +
[\partial_a + i e {\cal A}_a] \Gamma^a\psi \Bigr),\nonumber\\
&& a = 1,2; \, I, J = 1,2,3 \label{Green}
\end{eqnarray}

When the model is considered in lattice regularization, the values of momenta
belong to the Brillouin zone. The lattice regularization
contains mass parameter $m$ (for the details see \cite{Itep2012}). It has to
remain nonzero for the numerical algorithm to stay at work.  Physical results
are to be obtained when the extrapolation to $m=0$ is made.

%\subsection{Momentum space topology and the fermion excitations} \label{Moment}

%\subsection{Topological invariant in momentum space and zeros of $f_a, a = 1,2,3$}

%\subsection{$4D$ notations}

\begin{figure}
 \includegraphics[width=8cm]{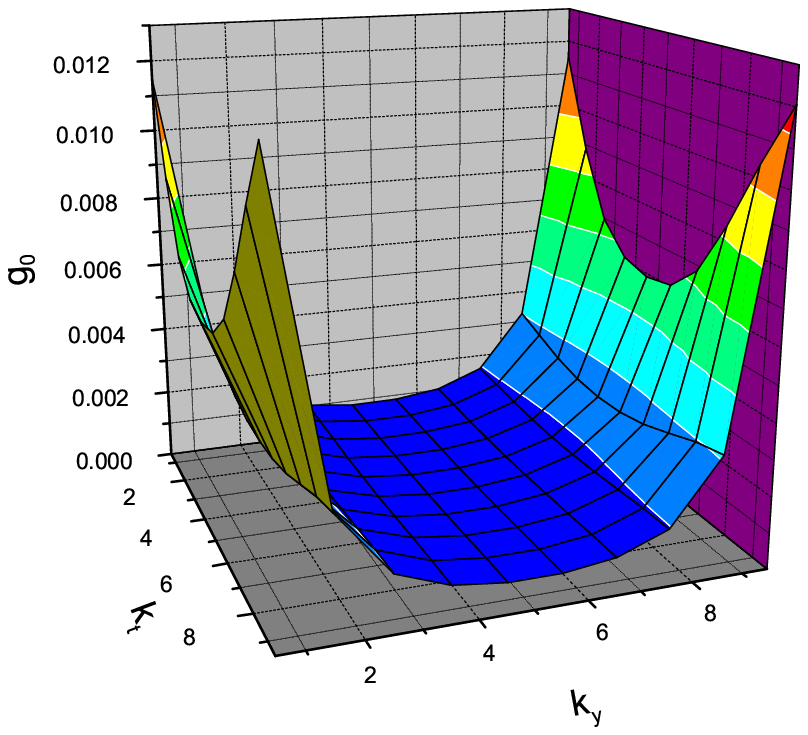}
 \includegraphics[width=8cm]{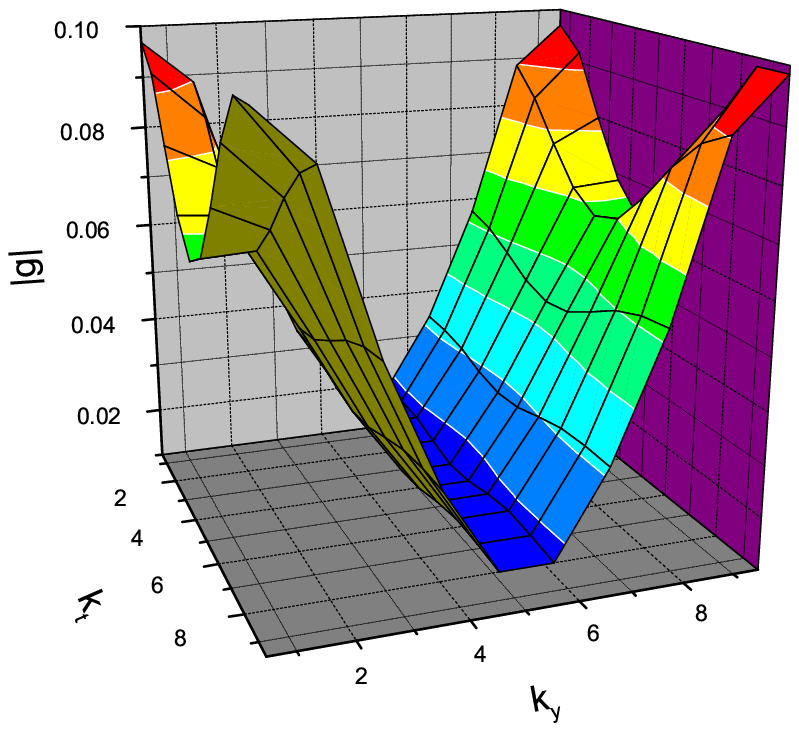} \\
 \caption{The values of $g_0$ (left figure) and $|g|$ (right figure) at $k_1=0$ in the semi - metal phase ($\beta = 0.2$).
  The lattice size is $20^3$. Error bars are within $2\%$ of the considered quantities.  We denote here $k_t=k_4, k_x=k_1, k_y=k_2$. }
 \label{g_val_semimetal}
\end{figure}

In the lattice regularized model with staggered fermions the single Grassman variable $\Psi$ is attached to  the sites \cite{Montvay}.
In terms of $\Psi$ the free fermion action has the form:
\begin{equation}
S = \sum_x\Bigl( m\, \bar{\Psi}_x\Psi_x + \frac{1}{2}\sum_{i = 1,...,4}
[\bar{\Psi}_x \alpha_{xi} \Psi_{x + \hat{i}} - \bar{\Psi}_{x+\hat{i}}
\alpha_{xi} \Psi_{x }]\Bigr),\quad \alpha_{xi} = (-1)^{x_1 + ... + x_{i-1}}
\end{equation}

In order to return to the original spinor and flavor indices of the spinors the lattice is considered with even number of lattice spacings in
each direction \cite{Montvay}. Let us subdivide this lattice into the blocks consisted of elementary cubes. Each block has $2^3$ sites (two
lattice sites in each direction). We denote the coordinates of the blocks by $y_i$. Therefore, the coordinates of the lattice sites are $x_i = 2
y_i + \eta_i, \eta_i = 0,1$. We define the new fields
\begin{equation}
[\Phi_y]^{\alpha}_a = \frac{1}{8}\sum_{\eta} [\Gamma_1^{\eta_1}
\Gamma_2^{\eta_2} \Gamma_4^{\eta_4}]^{\alpha}_a \Psi_{2 y + \eta}\label{PHI}
\end{equation}
Here index $\alpha = 1, ..., 4$ is the spinor index while $a = 1, ..., 4$ is
the flavor index. Matrices $\Phi$ have $4\times 4$ components. But not all of these components are independent. Eq. (\ref{PHI}) leads to the constraint
\begin{equation}
\Gamma_3 \Gamma_5 \Phi_y \Gamma_5 \Gamma_3  = \Phi_y
\end{equation}
This constraint reduces the number of flavors from $4$ to $2$. The free propagator of $\Phi$   in momentum representation (of the blocked lattice) has the form (see \cite{Montvay} and also \cite{Z2011,Itep2012}):
\begin{eqnarray}
 &&\tilde{\bf G} =   \Bigl( \sum_a \Gamma_a \frac{1}{2}{\rm sin}\, k_a - i (m -
\sum_a \frac{1}{2}(1 - {\rm cos}\, k_a)\Gamma_5 \otimes T_5 T_a) \Bigr)^{-1}\nonumber\\ & = & \frac{ \frac{1}{2}\sum_a \Gamma_a {\rm sin}\, k_a
+ i (m - \frac{1}{2}\sum_a (1 - {\rm cos}\, k_a)\Gamma_5 \otimes T_5 T_a)
}{32[\sum_a \frac{1}{2}(1-{\rm cos}\, k_a) + m ^2]}\label{PROP}
\end{eqnarray}
Here $T_i = \Gamma_i^T$ acts on the flavor indices while $\Gamma$ matrices act
on the Dirac indices. Momenta $k$ are $ k_1 = \frac{2\pi K_1}{N/2}\, \quad k_2
= \frac{2\pi K_2}{N/2} \, \quad k_4 = \frac{2\pi K_4 + \pi}{N/2}, \quad
K_1,K_2,K_4 \in Z $; the lattice size is  $N^3$.
At the end of the calculation one must set $m =
0$.  The terms proportional to $(1 - {\rm cos}\, k_a)\sim k_a^2 \sim a^2$ disappear in the continuum limit (the other terms are proportional to $\sim a$; here $a$ is the lattice spacing).

We suppose that when the interaction is switched on the form of the Green function is the same:
\begin{equation}
 \tilde{\bf G} = g_1 \Gamma_1 + g_2 \Gamma_2 + g_3 \Gamma_4 +
i g_0 +i  g^a_5 \Gamma_5 \otimes T_5 T_a.\label{GGAMMA}
\end{equation}
The terms proportional to $g^a_5$ are expected to be negligible in the continuum limit similar to the corresponding terms without the Coulomb interactions. The values of $g_a, a=1,2,3$ can be calculated as
\begin{eqnarray}
g_a (k)& = & \frac{i}{16 N_1^2 N_2^2 N_t^2} \sum_{y,z} e^{i k (z -
y)}\sum_{\eta,\eta^{\prime} } (-1)^{\eta_1 +...+\eta_{a-1}}\nonumber
\\ &&\delta(\eta^{\prime}_i-[\eta_i + \delta_{ia}]{\rm mod} \, 2)\langle G(2 y + \eta, 2 z + \eta^{\prime})\rangle
\end{eqnarray}
Here $\langle G(2 y + \eta, 2 z + \eta^{\prime})\rangle$ is the staggered
fermion  propagator in the external field averaged over the
configurations of the $U(1)$ gauge field and over the pseudofermion
configurations.

In addition we calculated the $g_0$ component of the Green function as follows:
\begin{eqnarray}
g_0 (k)& = & \frac{i}{16 N_1^2 N_2^2 N_t^2} \sum_{y,z} e^{i k (z -
y)}\sum_{\eta,\eta^{\prime} } \delta(\eta^{\prime}_i-[\eta_i + \delta_{ia}]{\rm
mod} \, 2)\langle G(2 y + \eta, 2 z + \eta^{\prime})\rangle
\end{eqnarray}

\begin{figure}
 \includegraphics[width=8cm]{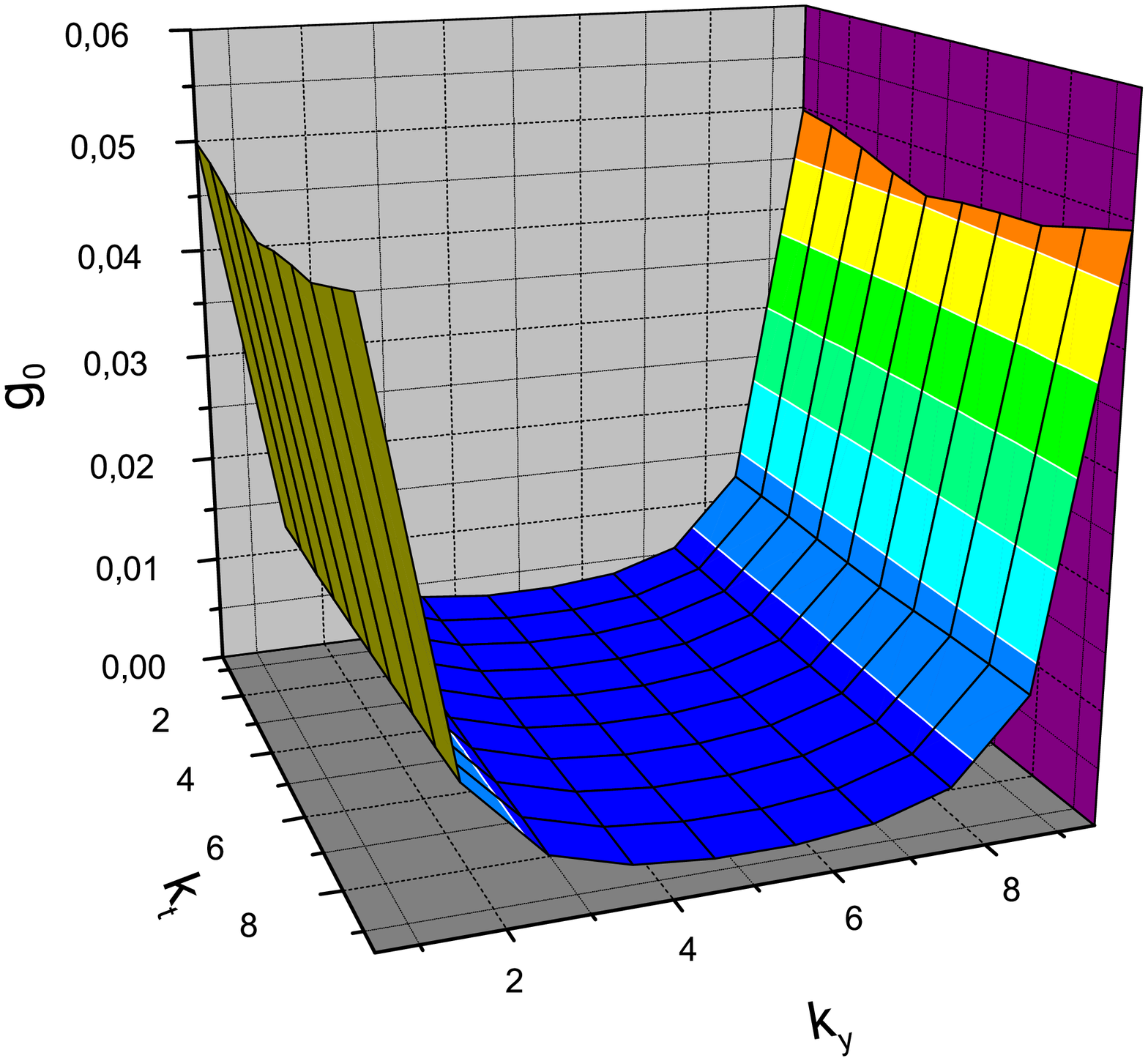}
 \includegraphics[width=8cm]{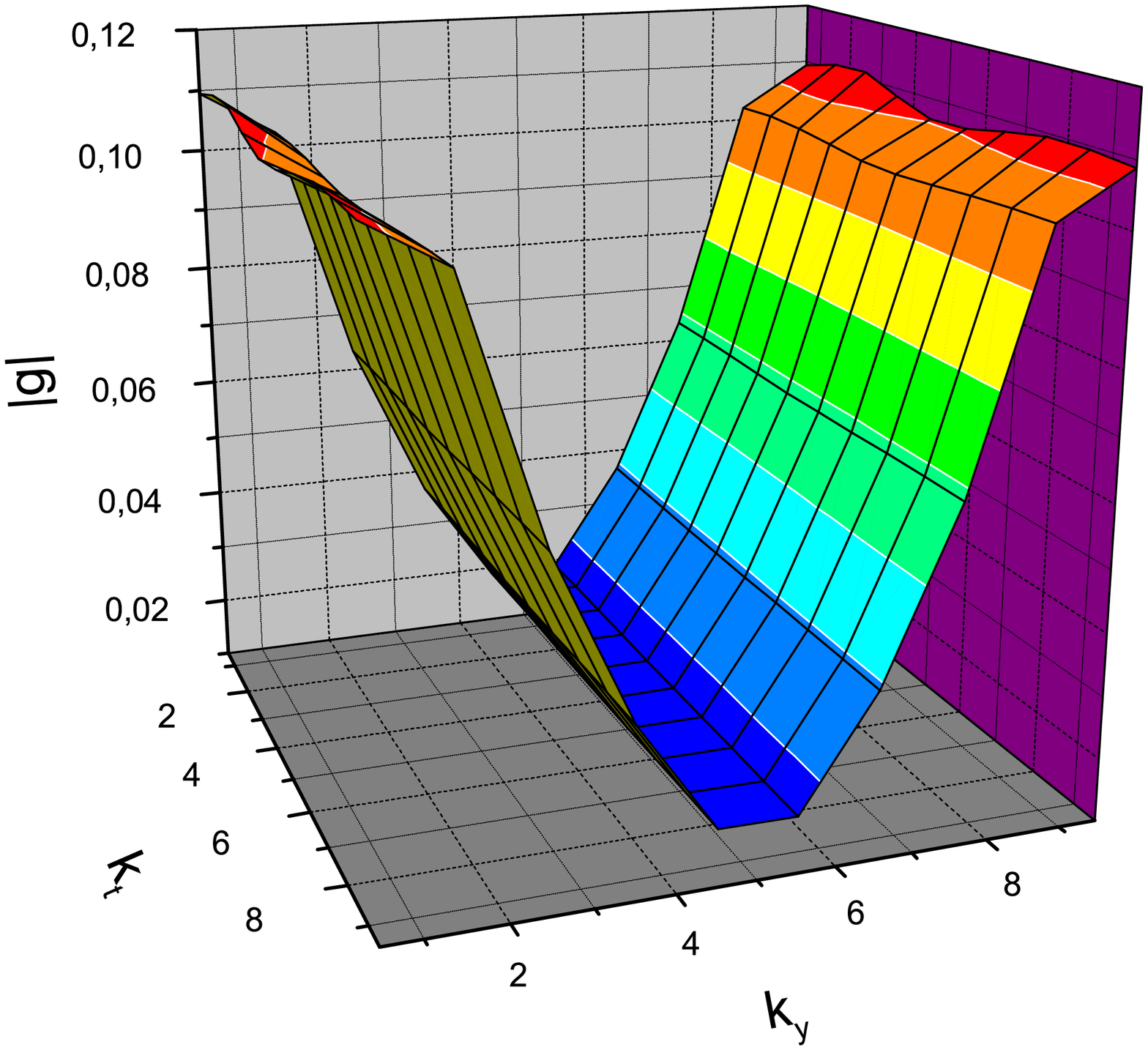} \\
 \caption{The values of $g_0$ (left figure) and $|g|$ (right figure) at $k_1=0$  close to the phase transition ($\beta = 0.08$).
 The lattice size is $20^3$. Error bars are within $2\%$ of the considered quantities.  We denote  $k_t=k_4, k_x=k_1, k_y=k_2$.}
 \label{g_val_pt}
\end{figure}

\section{Numerical results}
\label{SectNum}

%\subsection{The values of the Green functions.}

We simulate the model at $m = 0.01$. We collected enough statistics to
calculate the Green functions over all momentum space at $\beta = 0.05, 0.07,
0.08, 0.09, 0.1, 0.2$ on the lattices $10^3$ and $20^3$. On the smaller lattice
the Green functions are calculated using the direct inversion of matrices. On
the larger lattice of size $20^3$ we calculated the Green function using the
stochastic estimators (for the description of the method see \cite{Itep2012}).
According to \cite{Itep2012,Timo} the values $\beta = 0.05,0.07,0.08$ belong to
the insulator phase while the values $\beta = 0.09,0.1,0.2$ belong to the semi
- metal phase. We do not observe any qualitative dependence of the results on
the lattice size.

We  analyze the data on the values of the Green functions and have found that there is the essential excess of $|g| = \sqrt{g_1^2 + g_2^2 +
g_3^2}$ at $k_a \sim 0 \, (a = 1,2,3)$ over the average value within the momentum space lattice for $\beta = 0.2$. The dependence of this
quantity on momentum is represented in Fig.~\ref{g_val_semimetal}. On this figure we represent the values of $|g| = \sqrt{g_1^2 + g_2^2 +
g_3^2}$ and $g_0$ attached to the points of the dual lattice. The value at the point of the dual lattice is obtained via the averaging over the
vertices of the corresponding cube of the original lattice. Namely, we plot the values $|g| = \sqrt{\frac{1}{24}\sum_{a,v}g_{a}(v)^2}$ and $g_0
= \frac{1}{8}\sum_{v}g_{0}(v)$, where the sum is over the vertices $v$ of the given cube and over the components $a = 1,2,3$.
 On this figure the four peaks represent the single one due to the periodic boundary conditions.

\begin{figure}
 \includegraphics[width=8cm]{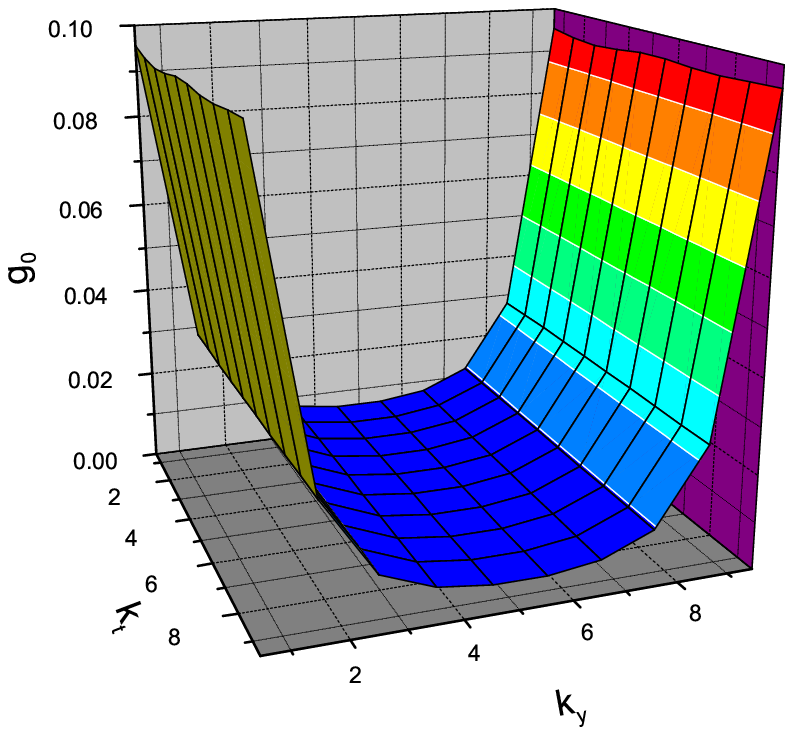}
 \includegraphics[width=8cm]{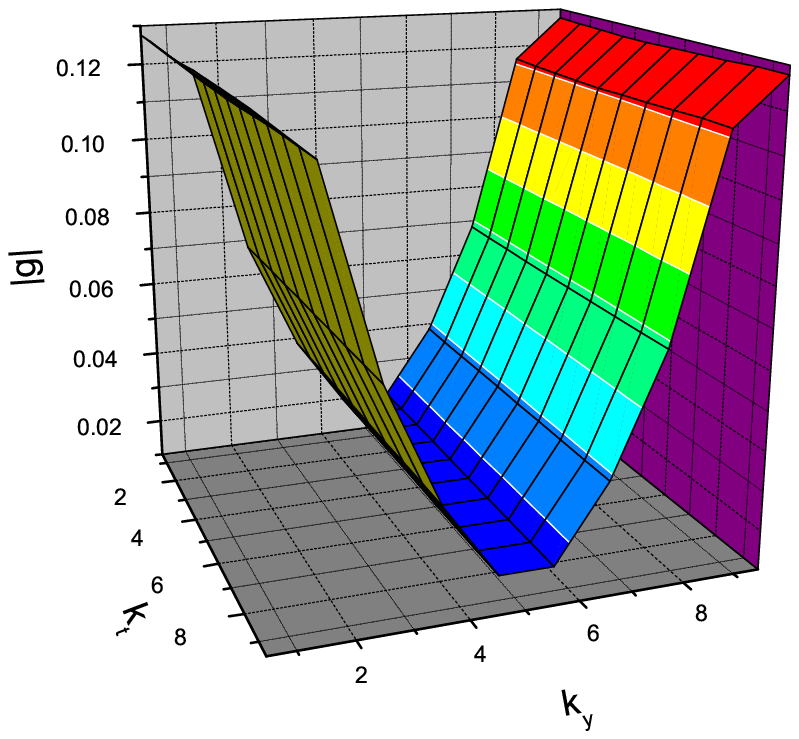} \\
 \caption{The values of $g_0$ (left figure) and $|g|$ (right figure) at $k_1=0$ in the insulator phase ($\beta = 0.05$). The lattice size is $20^3$. Error bars are within $2\%$ of the considered quantities. We denote here $k_t=k_4, k_x=k_1, k_y=k_2$.}
 \label{g_val_insulator}
\end{figure}

We observe that deep in the insulator phase (at $\beta = 0.05$)  the Green
function practically does not depend on $k_4$. Moreover, $g_3$ is negligible
compared to $g_1$ and $g_2$. This means that different time spices correlate
with each other very weakly, and that the system is described by the effective
$2D$ model rather than by the $2+1$ dimensional model. The dependence of the
Green function on $k_1, k_2$ demonstrates an essential excess of $|g| = \sqrt{
g_1^2 + g_2^2}$ at $k_1 = k_2 = 0$ over the rest of the momentum space lattice.
The value of $g_0$ at $k_1 = k_2 = 0$ in the insulator phase is essentially
larger than the value of $g_0$ in the semi - metal phase for $k_4 = k_1 = k_2
= 0$ (see Fig.~\ref{g_val_insulator}; again, the values are attached to the
points of the dual lattice and are averaged over the vertices of the
corresponding cubes).

Close to the phase transition ($\beta = 0.07, 0.08, 0.09, 0.1$) the maxima of
$|g|$ and $g_0$ as the functions of $k_4$ at $k_4 = 0, k_1 = 0, k_2 = 0$
are observed, in principle. However, the hights of these maxima are very small,
while $|g|$ and $g_0$ depend on $k_4$ very weakly (see Fig.~\ref{g_val_pt}).
That's why at $m=0.01$  we observe smooth transition between the two regimes in
the vicinity of the phase transition (its position is pointed out, for example, in
\cite{Itep2012,Timo}). The first regime corresponds to the effective $2D$
description of the theory approached deep in the insulator phase. The second
regime corresponds to the traditional semi - metal phase.

\section{Conclusions}
\label{SectConcl}

We have simulated the lattice regularized effective field model of graphene
monolayer with the Coulomb interactions taken into account. We calculate the
fermion Green function in momentum space. We consider several points on the phase diagram: deep in the the
semi - metal phase, close to the position of the phase transition pointed out
in
\cite{Timo,Itep2012,Drut:10:1,Hands:11:1,Lahde:09:1,Lahde:09:3,Lahde:11:1},
and deep in the insulator phase.

At $\beta = 0.05$ (deep in the insulator phase) the Green function practically
does not depend on $k_4$. This means that different time slices correlate
very weakly. Moreover, the values of $g_3(k) \sim \langle \bar{\psi}(k)
\Gamma_4 \psi(k) \rangle$ are negligible compared with $g_1,g_2$ for any values
of the momentum $k$
(where $g_1(k) \sim \langle \bar{\psi}(k) \Gamma_1 \psi(k) \rangle ,\, \quad
g_2 \sim \langle \bar{\psi}(k) \Gamma_2 \psi(k) \rangle$). This
means that deep in the insulator phase the energy of the fermionic
excitation tends to infinity.
Close to the position of the phase transition semi - metal -- insulator we
observe the intermediate behavior of the mentioned above quantities. Namely,
there are very small hights of the peaks of $|g| = \sqrt{g_1^2+g_2^2+g_3^2}$ and $g_0 \sim
\langle \bar{\psi}(k) \psi(k) \rangle$ as  functions of
$k_4$.

This is confirmed also by the consideration of the results for the current - current correlator
as a function of imaginary time represented in \cite{Itep2012} (Eq. (16)). Namely, in Fig. 3 of \cite{Itep2012} the spectral
density of this correlator is represented. It is clear from this Figure that in the insulator phase at $\epsilon < 4$ (i.e., for $\beta < 0.8$) the only maximum of the spectral density is at the frequencies $k_4 \sim 1/a$, where $a$ is the lattice spacing.
For $k_4 << 1/a$ the values of the spectral density are much less than at $k_4 \sim1/a$. Therefore, the correlation time extracted from this correlator is of the order of the lattice spacing (in the insulator phase) .

We consider the mentioned above results as an indication that at
the sufficiently small  values of the dielectric permittivity of the substrate the effective
low energy field model does not correspond to the original tight - binding
model. Therefore, it may have nothing to do with the reality. Most likely, here the
discreteness of the graphene honeycomb lattice becomes important for the
description of  the physical phenomena and the excitations that are not described
 by the effective field theory play an important role. Therefore the conclusion
of \cite{Araki,Araki:10:1,Timo,Itep2012,armour,Drut:10:1,Hands:11:1,Lahde:09:1,Lahde:09:3,Lahde:11:1,Itep2012} \footnote{This conclusion is made on the basis of the numerical investigation of the given effective theory.},
 that there is the insulator phase of the graphene monolayer,  seems to us questionable.

It is worth mentioning that we measure our quantities at fixed $m=0.01$ while
the phase transition was observed
in the behavior of the quantities extrapolated to $m = 0$.  In order to make definite
conclusions  it is necessary to repeat the calculations described in the present paper for different values of $m$ and to
 extrapolate the results to the value of $m$ equal to zero. Also the dependence
of the quantities on the lattice size has to be investigated. This should be a
content of the further investigation.

To conclude let us mention the recent work \cite{experiment}, where the experimantal results are presented with no sign of the insulator phase in graphene.

The authors kindly acknowledge comments of M.I.Katsnelson, and a private communication with G.E.Volovik, as well as the discussions with the members of the lattice ITEP group  M.I.Polikarpov,
P.V.Buividovich, V.I.Zakharov, O.V.Pavlovsky. The numerical simulations have
been performed using the facilities of the supercomputer centers of Moscow
University, Kurchatov  Institute, and ITEP.
 This work was
partly supported by RFBR grant 11-02-01227 and by the Russian
Ministry of Science and Education (program "Human Capital" and the contract No.
07.514.12.4028).

\end{document}